\documentclass[prd,superscriptaddress,amsfonts,amssymb,onecolumn,amsmath,showpacs]{revtex4-2}
\usepackage{bm}
\usepackage{amsfonts}
\usepackage{latexsym}
\usepackage[utf8]{inputenc}
\usepackage{graphicx}
\usepackage{amsmath}
\usepackage{palatino}
\usepackage{mathpazo}
\usepackage[british]{babel}
\usepackage{hhline}
\usepackage{multirow}
\usepackage{textcomp}
\usepackage{float}
\usepackage{booktabs}
\usepackage{dcolumn}
\usepackage{lipsum} 
\usepackage{mdframed}
\usepackage[]{mdframed}
\usepackage{hhline}
\usepackage{multirow}
\usepackage{ragged2e}
\usepackage{hyperref}
\hypersetup{colorlinks,citecolor=blue}
\hypersetup{colorlinks=true,linkcolor=red,filecolor=magenta,    urlcolor=cyan}
\usepackage{amsmath}
\usepackage{xcolor}
\usepackage{orcidlink}
\usepackage{epsfig}
\usepackage{caption}
\usepackage{subcaption}
\usepackage{commath}

\def\jnl@style{\it}
\def\aaref@jnl#1{{\jnl@style#1}}

\def\aaref@jnl#1{{\jnl@style#1}}

\def\aj{\aaref@jnl{AJ}}                   
\def\apj{\aaref@jnl{ApJ}}                 
\def\apjl{\aaref@jnl{ApJ}}                
\def\apjs{\aaref@jnl{ApJS}}               
\def\apss{\aaref@jnl{Ap\&SS}}             
\def\aap{\aaref@jnl{A\&A}}                
\def\aapr{\aaref@jnl{A\&A~Rev.}}          
\def\aaps{\aaref@jnl{A\&AS}}              
\def\mnras{\aaref@jnl{Mon.~Not.~Roy.~Astron.~Soc.}}             
\def\prd{\aaref@jnl{Phys.~Rev.~D}}        
\def\prc{\aaref@jnl{Phys.~Rev.~C}}  
\def\prl{\aaref@jnl{Phys.~Rev.~Lett.}}    
\def\qjras{\aaref@jnl{QJRAS}}             
\def\skytel{\aaref@jnl{S\&T}}             
\def\ssr{\aaref@jnl{Space~Sci.~Rev.}}     
\def\zap{\aaref@jnl{ZAp}}                 
\def\nat{\aaref@jnl{Nature}}              
\def\aplett{\aaref@jnl{Astrophys.~Lett.}} 
\def\apspr{\aaref@jnl{Astrophys.~Space~Phys.~Res.}} 
\def\physrep{\aaref@jnl{Phys.~Rep.}}      
\def\physscr{\aaref@jnl{Phys.~Scr}}       
\def\commat{\aaref@jnl{Comm.~Math.~Phys.}}              
\def\science{\aaref@jnl{Science}}               
\def\cqg{\aaref@jnl{Classical Quant.~Grav.}}            
\def\jpcs{\aaref@jnl{JPCS}}                                     
\def\ijmpd{\aaref@jnl{Int.~J.~Mod.~Phys.~D}}                    
\def\grg{\aaref@jnl{Gen.~Relat.~Gravit.}}               
\def\rpp{\aaref@jnl{Rep.~Prog.~Phys.}}          
\def\npa{\aaref@jnl{Nucl.~Phys.~A}}        
\def\lrr{\aaref@jnl{Living Rev.~Rel.}}                   
\def\jcap{\aaref@jnl{J.~Cosmology Astropart.~Phys.}}    
\def\rmp{\aaref@jnl{Rev.~Mod.~Phys.}}   
\def\epjc{\aaref@jnl{Eur.~Phys.~J.~C}} 
\def\plb{\aaref@jnl{~Phy.~Lett.~B}} 
\def\mpla{\aaref@jnl{Mod.~Phy.~Lett.~A}} 
\def\arxiv{\aaref@jnl{arxiv.org}}

\allowdisplaybreaks[1]

\begin{document}

\title{Constraining $f(T,B,T_G,B_G)$ gravity by dynamical system analysis}

\author{S. A. Kadam\orcidlink{0000-0002-2799-7870}}
\email{k.siddheshwar47@gmail.com}
\affiliation{Department of Mathematics, Birla Institute of Technology and Science-Pilani, Hyderabad Campus, Hyderabad-500078, India}

\author{B. Mishra\orcidlink{0000-0001-5527-3565}}
\email{bivu@hyderabad.bits-pilani.ac.in}
\affiliation{Department of Mathematics, Birla Institute of Technology and Science-Pilani, Hyderabad Campus, Hyderabad-500078, India}


\begin{abstract} {\textbf{Abstract:}} 
The evolutionary behavior of the Universe has been analysed through the dynamical system analysis in $f(T,B,T_G,B_G)$ gravity, where $T$, $B$, $T_G$, and $B_G$ respectively represent torsion, boundary term, teleparallel Gauss-Bonnet term and Gauss-Bonnet boundary term. We use the transformation, $f(T,B,T_G,B_G)=-T+\mathcal{F}(T, B, T_G, B_G)$ in order to obtain the deviation from the Teleparallel Equivalent of General Relativity (TEGR). Two cosmological models pertaining to the functional form of $\mathcal{F}(T, B, T_G, B_G)$ have been studied. The well motivated forms are: (i) $\mathcal{F}(T, B, T_G, B_G) = f_{0} T^{m} B^{n}T_{G}^{k}$ and (ii) $\mathcal{F}(T, B, T_G, B_G)=b_{0} B + g_{0} T_{G}^{k} $. The evolutionary phases of the Universe have been identified through the detailed analysis of the critical points. Further, with the eigenvalues and phase space diagrams, the stability and attractor nature of the accelerating solution have been explored. The evolution plots have been analyzed for the corresponding cosmology and compatibility with the present observed value of standard density parameters have been shown.

\end{abstract}

\maketitle


\section{Introduction}
Understanding the accelerating behaviour of the Universe is one of the most crucial challenges in modern cosmology. A significant study at the end of the nineteenth century remarked that we could attain high-precision distance measurements of type Ia supernovae (SNeIa) \cite{Riess:1998cb,Perlmutter:1998np} and revealed the accelerating expansion behaviour of the Universe. Subsequently, it has been confirmed through cosmic microwave background (CMB) anisotropy  \cite{Hinshaw:2013}, baryon acoustic oscillations and the large-scale clustering patterns of the galaxy \cite{BAO-2005}. Though the reason behind this behaviour is still to be established, it is believed that some unknown form of energy, known as dark energy (DE), is responsible for this strange behaviour. According to cosmological observations \cite{Riess:1998cb,Perlmutter:1998np,Aghanim:2018eyx,Hinshaw:2013,BAO-2005}, the present Universe is filled with DE \cite{Bayesianevidences2018}, dark matter \cite{Jarosik_2011}, and baryonic matter \cite{GELLMANN1964214}. Theoretically, to explain this behaviour, one can connect straightforward to the cosmological constant $\Lambda$, which is related to the $\Lambda$CDM model in standard cosmology; CDM denotes cold dark matter. It is well known that this model still suffers with the cosmological constant and coincidence problems \cite{peebles2003cosmological}. So, considering that General Relativity (GR) is valid at solar system scales and can be modified at extragalactic and cosmological scales, this theoretical issue may be resolved. Hence, GR can be modified to reflect the current behaviour of the Universe, thereby a new approach to understanding the DE and accelerating behavior of the Universe\cite{Cai:2015emx,Capozziello:2018qcp}.\\

Another equivalent approach of GR is the teleparallel equivalent of GR (TEGR), in which the gravitational effect would be provided by the torsion in place of curvature as in GR. The torsion would be generated from the tetrad vectors, and the connection that complies with this kind of geometry is known as a Weitzenböck connection \cite{Weitzenbock1923}. TEGR is equivalent to GR in terms of the field equations, but in terms of the background connections, the geometrical interpretation of TEGR is different from GR. The modification to TEGR can also be performed from the teleparallel point of view like the modification in GR, leading to several modified gravitational theories. One of the immediate modifications on TEGR is the  $f(T)$ gravity \cite{Bengochea:2008gz,Linder:2010py}, where $T$ is the torsion scalar, and it has been successful in addressing some of the key issues of the recent behaviour of the Universe. For more information one can refer, \cite{Finch:2018gkh,Farrugia:2016xcw,Bengochea:2008gz,Linder:2010py,Chen:2010va,Bahamonde:2019zea,Paliathanasis:2021nqa,Leon:2022oyy,Duchaniya:2022rqu,Kadam:2022lgq} and the review \cite{bahamonde:2021teleparallel,Cai:2015emx}. Again, the problem in $f(T)$ gravity is that it breaks local Lorentz invariance. Therefore, one can not fix any of the tetrad components \cite{Bahamonde:2016cul} due to a lack of local Lorentz symmetry. As a result, two tetrads corresponding to the same metric can produce different field equations. To overcome this problem, based on the principle of choosing a non-zero spin connection \cite{Weitzenbock1923,Krssak:2015oua}, the covariant formulation of $f(T)$ gravity has been developed  \cite{Krssak:2015oua,BLi2011}. To note, choosing the diagonal tetrads in Cartesian coordinates is important in the sense that they do not restrict the field equations, i.e., in both the covariant and non-covariant approaches, it yields similar results. Though the Lorentz invariance is lost, this approach is widely used in the literature\cite{Bengochea:2008gz,duchaniya2023dynamical,DUCHANIYA2024101402,LOHAKARE2023101164,Linder:2010py,Zubair:2015yma}. \\

In $f(T)$ gravity, dynamical equations are of second order, and in order to study the higher-order teleparallel theories, more number of terms to be introduced in the teleparallel Lagrangian. One among several such terms is the boundary term $B$, which depends on the derivative of the torsion tensor, and adding this term in the Lagrangian further generalised the $f(T)$ gravity to $f(T, B)$ gravity \cite{Bahamonde:2015zma}. It is noteworthy to mention, if one chooses the case $f(-T+B)$, the resulting theory becomes equivalent to $f(R)$ gravity, and this is the only case when Lorentz invariance can be achieved for the vanishing spin connection. Several aspects of $f(T, B)$ gravity have been studied, such as analysis of laws of thermodynamics and cosmological reconstruction methods \cite{Bahamonde:2016cul}, solar system tests \cite{farrugia2020gravitoelectromagnetism}, use of cosmological observational data sets \cite{Franco:2021,Escamilla-Rivera:2019ulu,briffa2023f}, bouncing cosmology \cite{caruana2020cosmological} and so on. Further, the cosmological dynamics have been analyzed using the dynamical system approach \cite{Franco:2021,Franco:2020lxx,Kadamdynamicalftb}. In addition, it is necessary to estimate the total degrees of freedom of a gravity theory by considering all possible geometric invariants at a given order. So, higher curvature corrections such as the Gauss-Bonnet term $G$ and the general function of it like $f(R, G)$ are explored in the curvature gravity \cite{Cognola_2006_73,glavan2020einstein}. Analogous to this, in the TEGR approach, the $f(T, T_G)$ gravity has been proposed \cite{Kofinas:2014owa}, where $T_G$ is the teleparallel equivalent of the Gauss-Bonnet term. This theory has been significant in describing important cosmological applications \cite{Kofinas:2014aka,Kofinas:2014daa,KADAM2024Aop,LOHAKARE2023101164,delaCruz-Dombriz:2017lvj,delaCruz-Dombriz:2018nvt,Capozziello:2016eaz,Kadam:2023}. Introducing Gauss-Bonnet boundary term ($B_G$) and teleparallel Gauss-Bonnet term ($T_G$) in $f(T,B)$ Lagrangian, the $f(T, B, T_G, B_G)$ gravity has been presented in \cite{bahamonde2016modified}. Both $T_G$ and $B_G$ are boundary terms in four dimensions, and these scalars contribute to the field equations only if the nonlinear forms are included in the action. This theory shows its generality by retrieving the higher order formalism in GR, $f(R, G)$ \cite{Santos2018} for the choice of $f = f(-T + B, -T_G + B_G)$ and the teleparallel Gauss-Bonnet gravity $f(T, T_G)$ for the particular choice of $f=f(T, T_G)$ \cite{Kofinas:2014aka,Kofinas:2014owa}. \\

It has been complicated to solve the field equations of higher-order gravity theory in order to study the cosmological aspects of the Universe. So, to obtain and analyse different epochs of the evolution of Universe, the dynamical system analysis approach can be considered \cite{KADAM2024Aop,Narawade:2022a}. Using this approach, the phase space and stability analysis can reveal the global features of a given cosmological scenario, which is independent of initial conditions. For more details, one can refer to the review articles \cite{bohmer2017dynamicalsystem,BAHAMONDEreview2018}. The dynamical system approach has been used to study both teleparallel and curvature Gauss-Bonnet modification \cite{Santos2018,Kofinas:2014aka,KADAM2024Aop,Lohakare_2303.14575}. Our motivation is to analyse different epochs of the Universe and their corresponding cosmological behaviour in the more general $f(T, B, T_G, B_G)$ gravity. The paper is organized as follows: in sec-\ref{sec:intro_tg}, the field equations of $f(T, B, T_G, B_G)$ gravity have been presented. The dynamical system analysis has been developed in sec-\ref{Dynamicalanalysis}, and for well-motivated functional forms of $f(T, B, T_G, B_G)$, the detailed dynamical system analysis for two models has been presented in subsec-\ref{Model-I} and subsec-\ref{Model-II}. The results and conclusions have been summarised in Sec-\ref{Conclusion}.

\section{\texorpdfstring{$f(T,B,T_G,B_G)$}{}  Gravity Field Equations}\label{sec:intro_tg}
In TEGR formulation, the fundamental variables are the tetrad fields $e^a_{\mu}$ and inverse tetrad fields $E^{a}_{\mu}$\cite{Krssak:2018ywd,Cai:2015emx,Aldrovandi:2013wha}, which provides an orthonormal basis for the tangent space at every point of the space-time manifold. The tetrads and their inverse adhere to the orthogonality relations
\begin{align}
    e^{a}_{\ \ \mu} E_{b}^{\ \ \mu}=\delta^a_b\,,&  & e^{a}_{\ \ \mu} E_{a}^{\ \ \nu}=\delta^{\nu}_{\mu}\,.
\end{align}
The Greek and the Latin indices are respectively used to represent space-time and the tangent coordinates \cite{Cai:2015emx}. The relationship between the metric $g_{\mu\nu}$, the inverse metric $g^{\mu\nu}$ with the tetrads $e^a_{\mu}$ and its inverse $E^{a}_{\mu}$ can be expressed as,
\begin{align}\label{metric_tetrad_rel}
    e^{a}_{\mu} e^{b}_{\nu}\eta_{ab}=g_{\mu\nu}\,,& &  E_{a}^{\ \ \mu} E_{b}^{\ \ \nu}g_{\mu\nu}=\eta_{ab} \,.
\end{align}
The teleparallel Weitzenb\"{o}ck connection \cite{Weitzenbock1923, Krssak:2015oua} can be defined as,
\begin{equation}
    W_{\mu\,\,\,\nu}^{\,\,a} :=
\partial_{\mu} e^{a}_{\ \ \nu} \,.
\end{equation}
From which, the torsion tensor \cite{Hayashi:1979qx} can be defined as the skew-symmetric part of the Weitzenb\"{o}ck connection,
\begin{equation}
    T^{a}_{\ \ \mu\nu}:=W_{\mu\,\,\,\nu}^{\,\,a}-W_{\nu\,\,\,\mu}^{\,\,a}=\partial_{\mu}e^{a}_{\nu}-\partial_{\nu}e^{a}_{\mu}\,.
\end{equation}
We can also define the contortion tensor as,
\begin{equation}
    \frac{1}{2}\left(T_{\mu \ \ \nu}^{\ \ \sigma}+ T_{\nu\ \ \mu}^{\ \ \sigma} - T^{\sigma}_{\ \ \mu\nu}\right) :=K^{\sigma}_{\ \ \mu\nu}\,.
\end{equation}
The action of the gravitational theory has been framed with the torsion scalar $T$ recovered by contractions of the torsion
tensor\cite{Krssak:2018ywd,Cai:2015emx,Aldrovandi:2013wha,bahamonde:2021teleparallel}
\begin{equation}\label{eq:torsion_scalar_def}
   \frac{1}{4}T^{\alpha}_{\ \ \mu\nu}T_{\alpha}^{\ \ \mu\nu} + \frac{1}{2}T^{\alpha}_{\ \ \mu\nu}T^{\nu\mu}_{\ \ \  \  \alpha} - T^{\beta}_{\ \ \beta\mu}T_{\nu}^{\ \ \nu\mu}:=T\,.
\end{equation}
This can be directly related to the curvature Ricci scalar $R$, which has been computed with the Levi-Civita connection as  \cite{bahamonde:2021teleparallel,Bahamonde:2015zma},
\begin{equation}\label{LC_TG_conn}
    -T+\frac{2}{e}\partial_{}\mu (e T^{\mu})=-T + B=R\,.
\end{equation}

In Eq.~(\ref{LC_TG_conn}), the curvature scalar $R$ and torsion scalar $T$ differ by a boundary term $B$: hence, the variations with respect to tetrads give the Einstein field equations. Therefore, GR and TEGR produce the same equations of motion, giving rise to the same dynamics.\\

The Gauss-Bonnet term is an interesting scale invariant that contains a combination of contraction of Riemann tensor and its quadratic form \cite{Kofinas:2014daa,Kofinas:2014owa,Zubair:2015yma,delaCruz-Dombriz:2017lvj,delaCruz-Dombriz:2018nvt,bahamonde2019noether} as,
\begin{equation}
   {R}^{2} - 4{R}_{\mu\nu}{R}^{\mu\nu} + {R}_{\mu\nu\alpha\beta}{R}^{\mu\nu\alpha\beta} =G \,.
\end{equation}
With the curvature approach, the Gauss-Bonnet scalar $G$ is a boundary term in 4 dimensions, and the addition of $G$ to the action of GR does not affect the resulting equations of motion. However, with more than four dimensions, it affects the gravitational field equations \cite{bahamonde2016modified}. In the teleparallel setting, the Gauss-Bonnet scalar \cite{Kofinas:2014owa,Kofinas:2014aka,Kofinas:2014daa} can be derived as,
\begin{align}\label{eq:T_G_def}
    T_G &= \Big(K_{a \ \ e}^{\ \ i} K_{b}^{\ \ ej}K_{c \ \ f}^{\ \ k} K_{d}^{\ \ fl} - 2K_{a}^{\ \ ij} K_{b \ \ e}^{\ \ k} K_{c \ \ f}^{\ \ e} K_{d}^{\ \ fl}  + 2K_{a}^{\ \ ij}K_{b \ \ e}^{\ \ k}K_{f}^{\ \ el}K_{d \ \ c}^{\ \ f} + 2K_{a}^{\ \ ij}K_{b \ \ e}^{\ \ k} K_{c,d}^{\ \ \ \ el} \Big)\delta_{ijlk}^{abcd},
\end{align}
where $\delta_{ijlk}^{abcd}$ is the generalized Kronecker delta. It has been discovered that the Teleparallel Gauss-Bonnet term $T_G$ and boundary Gauss-Bonnet term ($B_G$), have the similar relationship to Eq.(\ref{LC_TG_conn}) \cite{Kofinas:2014aka,Kofinas:2014daa} as, 

\begin{equation}
     -T_G + B_G=G \,,
\end{equation}
where,
\begin{equation}
    \frac{1}{e}\delta^{abcd}_{ijkl}\partial_{a} \left[K_{b}^{\ \ ij}\left(K_{c\ \ ,d}^{\ \ kl} + K_{d \ \ c}^{\ \ m}K_{m}^{\ \ kl}\right)\right]=B_G  \,.
\end{equation}
To note, as long as non-linear elements of these scalars are considered in action, the scalars $T_G$ and $B_G$ do not contribute to the field equations. The following action of the modified gravity helps to retrace the curvature and teleparallel Gauss-Bonnet gravity \cite{bahamonde:2021teleparallel,bahamonde2016modified},
\begin{equation}\label{f_T_G_ext_Lagran}
    \mathcal{S}_{f(T,B,T_G,B_G)}^{} =  \frac{1}{2\kappa^2}\int \mathrm{d}^4 x\; e\,f(T,B,T_G,B_G) + \int \mathrm{d}^4 x\; e\mathcal{L}_{\text{m}}\,,
\end{equation}
where $\kappa^2= 8\pi G$ and $\mathcal{L}_{\text{m}}$ be the matter Lagrangian. For the flat, homogeneous, and isotropic cosmological background, the tetrad can be defined as, 
\begin{equation}\label{flrw_tetrad}
    e^{A}_{\ \ \mu}=\textrm{diag}(1,a(t),a(t),a(t))\,,
\end{equation}
where $a(t)$ be the scale factor and can reproduce the spatially flat homogeneous and isotropic FLRW metric as,
\begin{equation}
    ds^2 = -dt^2+a(t)^2(dx^2+dy^2+dz^2)\,.
\end{equation}
It is worth mentioning that for the FLRW metric with vanishing spatial curvature, the higher order Gauss-Bonnet boundary term $B_G$ vanishes and hence $f(T, B, T_G, B_G)$ form reduces to $f(T, B, T_G)$ \cite{bahamonde2019noether}. In this setting,
\begin{equation}\label{T,T_G, B}
    T =6H^2\,,\quad T_G = -24H^2\left(\dot{H}+H^2\right) \,,\quad B=6(3H^2 + \dot{H})\,,
\end{equation}
where $H$ be the Hubble parameter and $H=\frac{\dot{a}}{a}$. Now the field equations of $f=-T+\mathcal{F}(T, B, T_G, B_G)$ gravity \cite{bahamonde2016modified,bahamonde2019noether,bahamonde:2021teleparallel} can be obtained as,
\begin{align}
   f +6 H \dot{f}_B - 2T f_{T} -4 T H \dot{f}_{T_G} -B f_{B} -T_{G} f_{T_G}  = 2 \kappa^2 \rho, \label{1stFE}\\ 
   f-4H\dot{f}_{T}-8H^2 \ddot{f}_{T_G} -B f_{B} - \frac{2B f_{T}}{3} +\frac{2 T_G \dot{f}_{T_G}}{3H}-T_G f_{T_G} +2 \ddot{f}_{B}= -2\kappa^2 p\label{2ndFE},
\end{align}
where $\rho$ and $p$ respectively denote energy density and pressure for matter and radiation. To find the deviation from TEGR, we use the transformation $f=-T+\mathcal{F}(T, B, T_G, B_G)$, which in turn provides the pressure and energy density expression for the DE phase. So, we express the following, 
\begin{align}
3H^2=\kappa^2\left(\rho+\rho_{DE}\right),\label{FridE1}\\
3H^2+2\dot{H}=-\kappa^2\left(p+p_{DE}\right)\label{FridE2}.
\end{align}
Comparing Eqs. (\ref{1stFE}-\ref{2ndFE}) and Eqs.(\ref{FridE1}-\ref{FridE2}), one can retrieve the field equations required to analyze the dynamics of DE as,

\begin{align}
    \frac{-1}{2\kappa^2}\left(\mathcal{F}+6H\dot{\mathcal{F}}_B-2T \mathcal{F}_{T}-4TH\dot{\mathcal{F}}_{T_G} -B \mathcal{F}_B-T_G \mathcal{F}_{T_G}\right)&=\rho_{DE}\,,\label{fede1}\\
    \frac{1}{2\kappa^2}\left(\mathcal{F}-4H\dot{\mathcal{F}}_T-8H^2\ddot{\mathcal{F}}_{T_G} -B \mathcal{F}_B -\frac{2B} {3} \mathcal{F}_{T} +\frac{2T_G}{3H} \dot{\mathcal{F}_{T_G}}-T_G\mathcal{F}_{T_G}+2\ddot{\mathcal{F}_{B}}\right)&=p_{DE}\,\label{fede2}
\end{align}
where $\rho_{DE}$ and $p_{DE}$ respectively denotes pressure and energy density for DE phase. Subsequently, it satisfies the continuity equation,
\begin{align}
\dot{\rho}_{DE}+3H\left(\rho_{DE}+p_{DE}\right)=0,
\end{align}
which also holds for the expressions of pressure and energy density of matter and radiation phases. 
\section{Dynamical system in \texorpdfstring{$f(T,B,T_G,B_G)$}{} 
Gravity}\label{Dynamicalanalysis}
The field equations obtained in Eqs. (\ref{FridE1}-\ref{FridE2}) are highly non-linear, and to analyse this, the dynamical system approach can be applied, which can be applied to a wide range of space-time \cite{paliathanasis2021epjp}. Applying the dynamical system approach, the evolution equations are reduced to an ordinary differential equation, which describes a self-consistent phase space.
This allows for an initial analysis of these theories and suggests what kinds of models should be examined or identifying possible observational evidence \cite{Franco:2021,Franco:2021,Escamilla-Rivera:2019ulu,briffa2023f}. Two boundary terms, such as $B$ and $T_G$, are included in the formalism, and the other boundary term vanishes because of the space-time chosen.
Now, we have defined the dynamical variables as, 
\begin{align}
X=\mathcal{F}_{T_G} H^2 \,,\quad Y=\dot{\mathcal{F}}_{T_G}H \,,\quad Z=\frac{\dot{H}}{H^2} \,,\quad 
V=\frac{\kappa^2 \rho_r}{3H^2}\,,\quad
W=-\frac{\mathcal{F}}{6H^2}\,,\quad  \Psi=\mathcal{F}_B \,,\quad \Theta=\mathcal{F}^{'}_{B}.\label{generaldynamical variables}
\end{align}
Moreover, for the dynamical system to be autonomous, we define the additional dynamical variable $\lambda=\frac{\ddot{H}}{H^3}$\cite{Odintsov:2018,Franco:2021,KADAM2024Aop}. The standard density parameters expressions for matter $(\Omega_m)$, radiation $(\Omega_r)$ and DE $(\Omega_{DE})$ phase are respectively,
\begin{align}
\Omega_{m}=\frac{\kappa^2 \rho_{m}}{3H^2}, \quad \Omega_{r}=\frac{\kappa^2 \rho_{r}}{3H^2}, \quad \Omega_{DE}=\frac{\kappa^2 \rho_{DE}}{3H^2},
\end{align}
which satisfies the constrained equation,
\begin{align}
\Omega_{m}+\Omega_{r}+\Omega_{DE}=1.
\end{align}
Subsequently Eq.(\ref{FridE1}) can be written in terms of dynamical variables as,
\begin{align}
1=\Omega_{m}+\Omega_{r}+W-\Theta+2\mathcal{F}_{T}+4Y+\left(3+Z\right)\Psi-4ZX-4X,
\end{align}
where,
\begin{align}
    \Omega_{DE}=W-\Theta+2\mathcal{F}_{T}+4Y+(3+Z)\Psi-4ZX-4X.
\end{align}
Now, the general form of the autonomous dynamical system in this formalism can be formed by considering differentiation of the dimensionless variables with respect to $N=ln(a)$ as, 
 
\begin{align}
X^{'}&=2XZ+Y\,,\nonumber\\
Y^{'}&=YZ+\Gamma\,,\nonumber\\
Z^{'}&=\lambda-2Z^2\,,\nonumber\\
V^{'}&=-4V-2VZ\,,\nonumber\\
W^{'}&= -2Z\mathcal{F}_T+4\lambda x +16ZX+8Z^2X-6Z\Psi-\lambda \Psi -2ZW\,,\nonumber\\
\Psi^{'}&=\Theta \,,\nonumber\\
\Theta^{'}&= \Xi-\Theta Z\,.\nonumber\\
\label{generaldynamicalsystem}
\end{align}
Where ({$^{'}$}) represents the differentiation with respect to $N=lna$.
We denote $\Gamma=\ddot{\mathcal{F}}_{T_G}$ and $ \Xi=\frac{\ddot{\mathcal{F}}_B}{H^2}$ and using Eq.(\ref{FridE2}) and Eq. (\ref{fede2}), one can relate $\Gamma$ and $ \Xi$ in terms of the dynamical variables as,
\begin{align}
    \Gamma=\frac{1}{4} \left[ 3+2 Z +V-3W-2\mathcal{F}^{'}_T-9 \Psi-3 Z \Psi-\left(6+2Z\right)\mathcal{F}_T - 8 Z Y -8Y+12ZX+12X+\Xi\right]\,.\nonumber
\end{align}
Only $\mathcal{F}_T$ and $\Xi$ have remained to be converted into the dimensionless variable to form the autonomous dynamical system. To do this, we need to have some form of $\mathcal{F}(T, B, T_G, B_G)$. Hence, in the following subsections, we have considered two forms of $\mathcal{F}(T, B, T_G, B_G)$ to frame the cosmological models and understand the evolutionary behaviour of the Universe.


\subsection{\textbf{Model-IIIA}}\label{Model-I}
First, we consider the mixed power law form of $\mathcal{F}(T, B, T_G, B_G)$ \cite{bahamonde2019noether} as
\begin{align}
\mathcal{F}(T, B, T_G, B_G)=f_{0} T^{m} B^{n}T_{G}^{k} ,\label{firstmodel}
\end{align}
where $f_{0}, m, n, k$ are arbitrary constants. This form is capable of converting $\mathcal{F}_{T}$ into the dynamical variables as $\mathcal{F}_{T}=-m W$, and this will guarantee the autonomous dynamical system. The variables from Eq.(\ref{generaldynamical variables}) can show the  dependency relations as \begin{align}
X&=\mathcal{F}_{T_G} H^2=f_{0} k T_G^{k-1} T^{m} B^{n}H^{2}\nonumber\\
&=\frac{k\mathcal{F}H^2}{T_G}=\frac{-k\mathcal{F}}{24(\dot{H}+H^2)}=\frac{-k\mathcal{F}}{6H^2}\left(\frac{1}{4(\frac{\dot{H}}{H^2}+1)}\right)=\frac{Wk}{4(Z+1)}.
\end{align}
which implies,
\begin{align}
W=\frac{4X(Z+1)}{k}\,. \label{eqforz}
\end{align}
Some other dynamical variables from Eq.(\ref{generaldynamical variables}) can have direct relationship as,
\begin{align}
\Psi&=\frac{-4n}{k}\frac{(Z+1)X}{ (Z+3)}\,,\nonumber\\
\Theta&=-\frac{4 (n-1) n X (Z+1) (\lambda +6 Z)}{k (Z+3)^2}\,,\nonumber\\
Y&= X\left[2mZ+n\frac{(6Z+\lambda)}{(Z+3)}+\frac{(k-1)(\lambda+4Z+2Z^2)}{(1+Z)}\right]\,.\label{depedency relations}
\end{align}
The variables $W,\Psi, \Theta, Y$ can be written in terms of the variables $X, Z, \lambda$. It reduces the system into four independent variables: $X, Z, V$, and $\lambda$. In this case, the dynamical variable $\lambda$ is treated as a constant that plays a crucial role in identifying the evolutionary epochs of the Universe. The motivation behind considering $\lambda$ as a constant is that cosmological solutions can be retraced for its constant value; such as when  $\lambda=8$, it refers to the study of radiation-dominated epochs, $\lambda=0$ leads to de-Sitter Universe and $\lambda$ = $\frac{9}{2}$ refers to matter-domination era. With this setup, the dynamical system in Eq. (\ref{generaldynamicalsystem}) can be written as,
\begin{align}
X^{'}&=X \left(2 m Z+\frac{n (\lambda +6 Z)}{Z+3}+\frac{(k-1) (\lambda +2 Z (Z+2))}{Z+1}+2 Z\right)\,,\nonumber\\
Z^{'}&=\lambda -2 Z^2,\nonumber\\
V^{'}&=-2 V(Z+2).\label{DynamicalSystem1stmodel}
\end{align}

The density parameters can be presented as,
\begin{align}
\Omega_{DE}&=-\Theta -2 m W+3 \Psi +W-4 X Z-4 X+4 Y+\Psi  Z\,,  \\
\Omega_{m}&=1+\Theta +2 m W-3 \Psi -V-W+4 X Z+4 X-4 Y-\Psi  Z\,.
\end{align}
Using the dependency relations in Eq.(\ref{depedency relations}), the density parameters become,
\begin{align}
    \Omega_{DE}&=\frac{4 X}{k}\left[\frac{n^2 (Z+1) (\lambda +6 Z)}{(Z+3)^2}-\frac{n (\lambda +\lambda  (Z-k (Z+3))+Z (-6 k (Z+3)+Z (Z+13)+21)+9)}{(Z+3)^2}\right]\nonumber\\
    &\quad \quad+\frac{4X}{k}\left[2 m ((k-1) Z-1)+\frac{(k-1) (\lambda  k+(2 k-1) Z (Z+2)-1)}{Z+1}\right]\,,\\
 \Omega_{m}&=-V+\left[\frac{(Z+1) \left(k (Z+3)^2\right)-4 X \left(n^2 (Z+1)^2 (\lambda +6 Z)-n \sigma_1 (Z+1)+(k-1) \sigma_3 (Z+3)^2+\sigma_2\right)}{(Z+1) \left(k (Z+3)^2\right)}\right]\,,
\end{align}
where,
\begin{align}\nonumber
    \sigma_1&=\lambda +\lambda  (Z-k (Z+3))+Z (-6 k (Z+3)+Z (Z+13)+21)
    +9\,,\\\nonumber
    \sigma_{2}&=2 m (Z+1) (Z+3)^2 ((k-1) Z-1)\,,\\\nonumber
   \sigma_{3}&=\lambda  k+(2 k-1) Z (Z+2)-1 \,.\\\nonumber
\end{align}
Further, as the dynamical variable $Z=\frac{\dot{H}}{H^2},$ the total EoS parameter($\omega_{tot}$) and the deceleration parameter ($q$) can be presented in the form of dynamical variables as,
\begin{align}
    \omega_{tot}&=-1-\frac{2 Z}{3} , \nonumber\\
    q&=-1-Z\,.
\end{align}
To obtain the critical points at different epochs, we consider $X^{'}=0$, $Z^{'}=0$, $V^{'}=0$. Here, we discuss the stability criteria of the critical points based on the nature of eigenvalues. In an $n$-dimensional system, if $n$ number of eigenvalues exist for each critical point, then the following classification to be followed \cite{Gonzalez-Espinoza:2020jss}: (i) if all eigenvalues are real and have a negative sign, then it is the stable node or attractor whereas for all positive sign, it represents unstable node; (ii) if all the eigenvalues are real and at least two of them are having opposite signs, then it is the saddle (unstable) point; (iii) if the eigenvalues are complex, then it can be further classified as (a) if all the eigenvalues have the negative real part, then it is stable focus node, (b) if all the eigenvalues have positive real parts, it represents an unstable focus node and (c) if at least two eigenvalues have real parts with opposite sign, it represents saddle focus. For Model-\ref{Model-I}, the coordinates and the values of $\omega_{tot}$, $q$ at each critical point are presented in Table \ref{modelIcriticalpoints}.  The eigenvalues and the stability conditions associated with each critical points are presented, and the conditions for which these critical points describe the accelerating phase of the Universe are presented in Table-\ref{modelIeigenvalues}. The standard density parameters for each critical point are given in Table-\ref{modelIdensityparametersm-1}.

\begin{table}[H]
    \centering 
    \begin{tabular}{|c |c |c |c|} 
    \hline\hline 
    \parbox[c][0.9cm]{1.3cm}{\textbf{Critical Points}
    }& \textbf{Co-ordinates} &  $\omega_{tot}$&  \textbf{$q$}\\ [0.5ex] 
    \hline\hline 
    \parbox[c][1.3cm]{1.3cm}{$A_1$ } &$\Big[\lambda =8, X=0, Z=-2, V= V_1\Big]$ & $\frac{1}{3}$ &  $1$\\
    \hline
    \parbox[c][1.3cm]{1.3cm}{$A_2$ } & $\Big[\lambda =8, X=X_2, Z=-2, V= V_2, k=\frac{-m-n+1}{2}\Big]$ & $\frac{1}{3}$&  $1$ \\
    \hline
   \parbox[c][1.3cm]{1.3cm}{$A_3$ } &  $\big[\lambda =\frac{9}{2}, X=0, Z=-\frac{3}{2}, V=0 \big]$ & $0$ &  $\frac{1}{2}$\\
   \hline
   \parbox[c][1.3cm]{1.3cm}{$A_4$} &  $\big[\lambda =2 Z_{4}^2, X=X_4, Z=Z_4, V=V_4, k=\frac{-m-n+1}{2}\big]$ & $-1-\frac{2Z_4}{3}$ &  $-1-Z_4$ \\
   \hline
 \parbox[c][1.3cm]{1.3cm}{$A_5$} &  $\big[\lambda =0, X=X_5, Z=0, V=0, \big]$ & $-1$ &  $-1$ \\
 \hline
    \end{tabular}
    \caption{Critical points and corresponding values of $\omega_{tot}$, $q$ for  Model-\ref{Model-I}.}
    \label{modelIcriticalpoints}
\end{table}
\begin{table}[H]
    \centering 
    \begin{tabular}{|c |c |c |c|} 
    \hline\hline 
    \parbox[c][0.9cm]{1.3cm}{\textbf{Critical Points}
    }& \textbf{Eigenvalues} & \textbf{Stability}& \textbf{Acceleration}\\ [0.5ex] 
    \hline\hline 
    \parbox[c][1.3cm]{1.3cm}{$A_1$ } & $\left[8,0,-4 \left(m+n+2 k-1\right)\right]$ & Unstable & Never \\
    \hline
    \parbox[c][1.3cm]{1.3cm}{$A_2$ } & $\left[8,0,0\right]$ & Unstable& Never \\
    \hline
   \parbox[c][1.3cm]{1.3cm}{$A_3$ } &  $\left[6,-1,-3 \left(m+n+2 k-1\right)\right]$ & Unstable & Never\\
   \hline
   \parbox[c][1.3cm]{1.3cm}{$A_4$ } &  $\left[0,-4 Z_4,-2 \left(Z_4+2\right)\right]$ &Stable for $Z_4>0$ &  $Z_4>-1$ \\
   \hline
   
   \parbox[c][1.3cm]{1.3cm}{$A_5$ } &  $\left[0,0 -4\right]$ & Nonhyperbolic & Always \\
 \hline
    \end{tabular}
    \caption{Eigenvalues corresponding to the critical points with stability and acceleration conditions for Model-\ref{Model-I}.}
    \label{modelIeigenvalues}
\end{table}
\begin{table}[H] 
    \centering 
    \begin{tabular}{|c |c |c |c|} 
    \hline\hline 
    \parbox[c][0.9cm]{1.3cm}{\textbf{Critical Points}
    }& $\Omega_r$ & $\Omega_m$&$\Omega_{DE}$\\ [0.5ex] 
    \hline\hline 
    \parbox[c][1.3cm]{1.3cm}{$A_1$ } & $V_1$ & $1-V_1$& $0$ \\
    \hline
    \parbox[c][1.3cm]{1.3cm}{$A_2$ } & $V_2$ & $1-V_2-\frac{4 X_2 \left(m-8 n^2+11 n-3\right)}{m+n-1}$ &$\frac{4 X_2 \left(m-8 n^2+11 n-3\right)}{m+n-1}$\\
    \hline
   \parbox[c][1.3cm]{1.3cm}{$A_3$ } &  $0$ & $1$ & $0$ \\
   \hline
   \parbox[c][1.3cm]{1.3cm}{$A_4$ } &  $0$ & $\frac{\left(n-1\right) \left(4 X_4 \left(Z_4 \left(4 n \left(Z_4+1\right)+Z_4+2\right)-3\right)+Z_4+3\right)-m \left(Z_4+3\right) \left(4 X_4 \left(Z_4+3\right)-1\right)}{\left(Z_4+3\right) \left(m+n-1\right)}$ &$\frac{4 X_4 \left(m \left(Z_4+3\right){}^2-\left(n-1\right) \left(Z_4 \left(4 n \left(Z_4+1\right)+Z_4+2\right)-3\right)\right)}{\left(Z_4+3\right) \left(m+n-1\right)}$ \\
   \hline
   \parbox[c][1.3cm]{1.3cm}{$A_5$ } &  $0$ &  $\frac{4 X_5 \left(k+2 m+n-1\right)}{k}+1$ &$4 X_5 \left(\frac{-2 m-n+1}{k}-1\right)$ \\
 \hline
    \end{tabular}
    \caption{Standard density parameters corresponding to the critical points for Model-\ref{Model-I}.}
    \label{modelIdensityparametersm-1}
\end{table}

\begin{figure}[H]
    \centering
    \includegraphics[width=65mm]{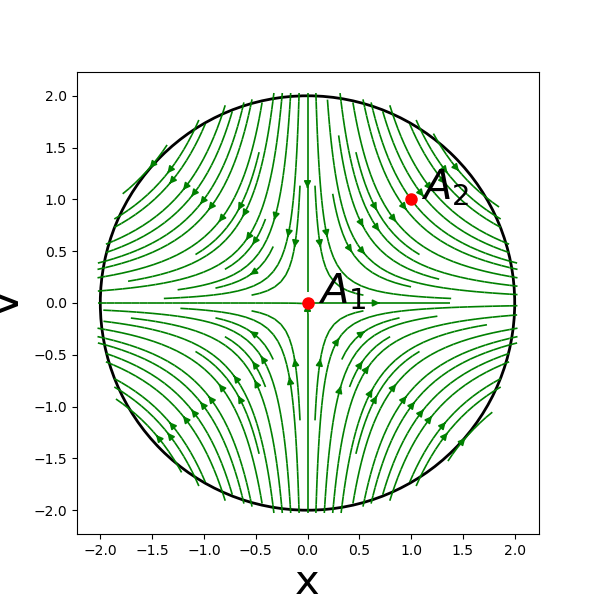}
    \includegraphics[width=65mm]{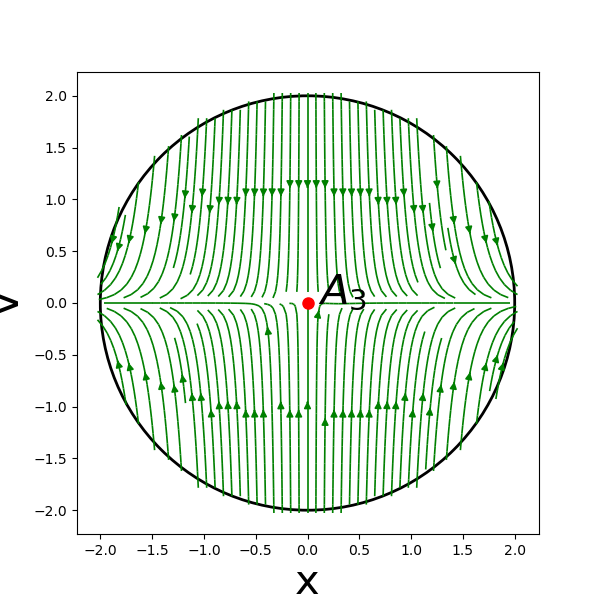}
    \includegraphics[width=65mm]{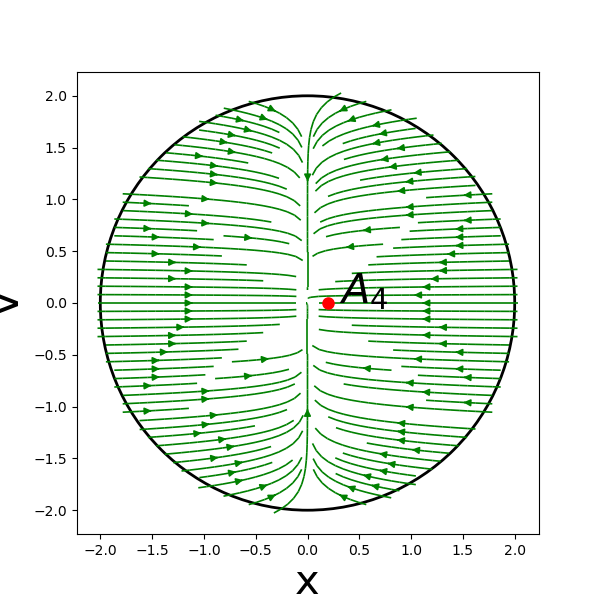}
    \includegraphics[width=65mm]{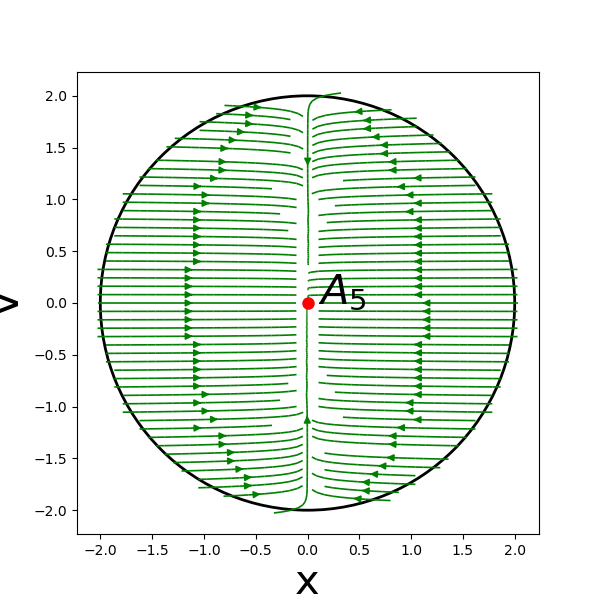}
    \caption{ $2D$ phase portrait for the dynamical system with $m=0.5, n=0.44, k=0.029$ ( Model-\ref{Model-I}). } \label{phasespacem1}
\end{figure}

A detailed discussion of each critical point has been given below according to the evolutionary history of the Universe.
\begin{itemize}
\item \textbf{Radiation-Dominated Critical Points:} Both critical points  $A_1$, $A_2$ are representing the early Universe radiation dominated era with $\omega_{tot}=\frac{1}{3}$ and $q=1$. Both are having eigenvalue eight as presented in Table \ref{modelIeigenvalues} with the positive signature are unstable. The value of $\Omega_{r}$ is dependent on the dimensionless variable $V$ refer Table \ref{modelIdensityparametersm-1} will represent a standard radiation-dominated era at $V_1=1$ for $A_1$ and for $A_2$ it is $V_2=1, X_2=0$. These critical points will not describe the current accelerating phase, and the phase space is plotted for the model parameter values $m=0.5, n=0.44, k=0.029$, which are similar values as discussed in \cite{KADAM2024Aop}. On analyzing the phase space trajectories presented in Fig. \ref{phasespacem1}, one can observe that the phase space trajectories are moving away at these critical points, confirming the unstable nature of these critical points. Moreover, in this case, the value of constant parameter $\lambda$ is 8.

\item \textbf{Matter-Dominated Critical Point:} This critical point $A_3$ represents standard matter-dominated era with $\Omega_{m}=1$ refer Table \ref{modelIdensityparametersm-1}. The values of $\omega_{tot}=0$ and $q=\frac{1}{2}$ at this critical point as presented in Table \ref{modelIcriticalpoints}. The value of the parameter $\lambda$ at this critical point is $\frac{9}{2}$. The phase space trajectory behavior can be analyzed using Fig. \ref{phasespacem1}. The phase space trajectories are moving away from this critical point, and hence, this critical point is showing unstable behavior. This critical point can be described in the important condition on model parameters $k=\frac{-m-n+1}{2}$, which is a similar condition that arises in \cite{KADAM2024Aop} for this particular model. This condition also plays a crucial role in obtaining a nontrivial Noether vector, as discussed in \cite{Capozziello:2016eaz,Bahamonde:2016grb} where particular forms of the model containing both $B$ and $ T_G$ have been discussed.

\item \textbf{DE-Dominated Critical Points:} The critical point $A_4$ at which we have $\omega_{tot}=-1-\frac{2 Z_4}{3}$ and $q=-1-Z_4$ as presented in Table \ref{modelIcriticalpoints}. This critical point will describe the acceleration of the Universe at $Z_4 \ge -1$ and show stability at $z_4\ge 0$ refer Table \ref{modelIeigenvalues}. The eigenvalues at this critical point contain zero, and the dimension of the set of eigenvalues is equal to a number of vanishing eigenvalues; hence, it is normally hyperbolic and shows stability at $Z_4 > 0$. This critical point will not represent a standard DE-dominated solution as the standard density parameter for matter contributes a small amount. The phase space trajectories attract at this critical point; hence, an attractor nature of this critical point can be analyzed from Fig. \ref{phasespacem1}.\\

The critical point $A_5$ represents a de-Sitter solution with $\omega_{tot}=q=-1$. This critical point is nonhyperbolic due to the presence of two zero eigenvalues. This critical point can describe the acceleration of the Universe and will describe the standard DE-dominated era at $X_5 = -\frac{k}{4 \left(k+2 m+n-1\right)}$. This critical point is also an attractor solution. The value of $\lambda$ at this critical point is 0, as expected. 
\end{itemize}

\begin{figure}[H]
    \centering
  \includegraphics[width=75mm]{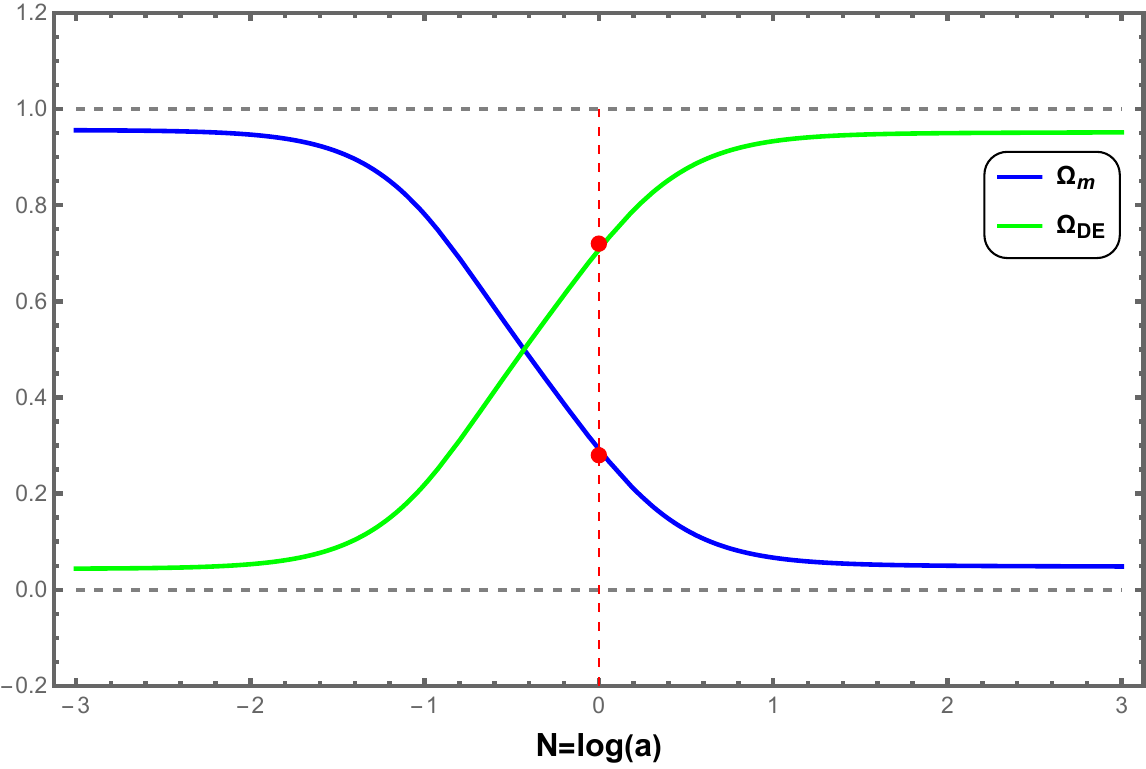}
    \includegraphics[width=75mm]{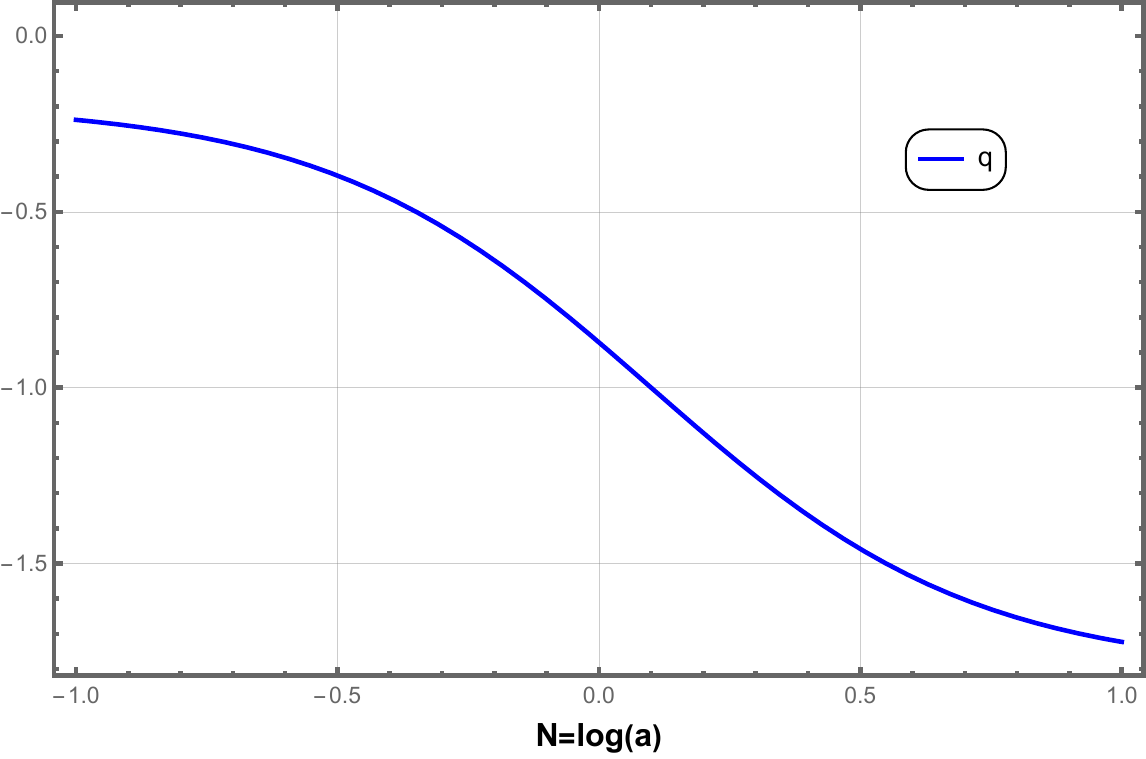}
    \caption{Behaviour of density parameters for matter and DE, $q$ in terms of the redshift for the dynamical system with $m=0.5, n=0.44, k=0.029$ ( Model-\ref{Model-I}). } \label{evolutionm1}
\end{figure}

From the evolution plots of standard density parameters presented in Fig.\ref{evolutionm1}, we observe that the matter density parameter was dominating the DE density parameter at early times and it is decreasing over time. At present, its values have been observed as $\Omega_{m}\approx 0.3$ which is compatible with the Planck observation results \cite{Aghanim:2018eyx}. The DE density parameter is increasing from early to late time and at present its value observed to be, $\Omega_{DE} \approx 0.7$ \cite{Kowalski_2008}. The deceleration parameter lies in the negative region and is capable of describing the accelerating behaviour of the Universe; the present value of $q \approx -0.810^{+0.1}_{-0.1}$ and is compatible with the result as in \cite{Capozziellomnras}.
\subsection{\textbf{Model-IIIB} }\label{Model-II}
The second form considered for $\mathcal{F}(T,B,T_G,B_G)$ is a particular form of $\mathcal{F}(T, B, T_G, B_G)= t_{0} T^m + b_{0}B^{n}+g_{0}T_{G}^{k}$, which has been successful in providing a viable cosmological model in Noether symmetry approach \cite{bahamonde2019noether}. For $k=0$ and $n=0$, the dynamical system analysis has been studied respectively in \cite{Franco:2020lxx} and \cite{KADAM2024Aop}. However, we are interested in analyzing a novel particular form in which the teleparallel boundary term $B$ and the Gauus-Bonnet invariant $T_G$ contribute. We define the second form as,

\begin{align}
\mathcal{F}(T, B, T_G, B_G)= b_{0}B+g_{0}T_{G}^{k} ,\label{secondmodel}
\end{align}
In this case, we have demonstrated that either one of $Y$ and $\lambda$ is dependent, not both at the same time. Here, we follow the approach in which the dynamical variable $Y$ is treated as independent and obtain the dependency relation for the variable $\lambda$; this will allow us to analyze all different phases of Universe evolution in a single phase space.  To obtain autonomous dynamical system we have $\mathcal{F}_T =0$ and the dependency relation as,
\begin{align}
W&=\frac{4 X}{k} \left( Z+1\right)-b_{0} k \left(3+Z\right) \,,\nonumber\\
\lambda&=\frac{Y (Z+1)}{(k-1) X}-2 Z (Z+2) \,.\label{dependencyrelationsmodel2}
\end{align}
Now, the autonomous dynamical system for the independent variables is,
\begin{align}
X^{'}&=2XZ+Y\,,\nonumber\\
Y^{'}&=\frac{1}{4} \left(\frac{12 (k-1) X (Z+1)}{k}+V-4 Y (Z+2)+2 Z+3\right)\,,\nonumber\\
Z^{'}&=\frac{(Z+1) (Y-4 (k-1) X Z)}{(k-1) X}\,,\nonumber\\
V^{'}&=-4V-2VZ\,.
\label{dynamicalsystemmodel2}
\end{align}

Also, with respect to the dynamical variables, the density parameters for DE and matter can be obtained respectively as,
\begin{align}
\Omega_{DE}&=4 Y-\frac{4 (k-1) X (Z+1)}{k}\,,  \\
\Omega_{m}&=1+\frac{4 (k-1) X (Z+1)}{k}-V-4 Y\,.
\end{align}

{\bf Note:} We represent the coordinates of the critical points with a small letter alphabet for this model. As in Model-\ref{Model-I}, the critical points for the system in Eq. (\ref{dynamicalsystemmodel2}) are presented in Table-\ref{modelIIcriticalpoints}. The eigenvalues and the stability conditions associated with each critical point are given in Table-\ref{modelIIeigenvalues} with the condition of acceleration. The standard density parameters for each critical point are given in Table-\ref{densityparametersm-2}.

\begin{table}[H]
    \centering 
    \begin{tabular}{|c |c |c| c|} 
    \hline 
    \parbox[c][0.9cm]{1.3cm}{\textbf{Critical Points}
    }& \textbf{Co-ordinates} &  $\omega_{tot}$ & $q$\\ [0.5ex] 
    \hline 

    \parbox[c][1.3cm]{1.3cm}{$B_1$ } &$\Big[ x=x_1\neq 0, y=4x_1, z=-2, v=1-12 x_1, k=\frac{1}{2}\Big]$ & $\frac{1}{3}$& $1$ \\
    \hline
    
    \parbox[c][1.3cm]{1.3cm}{$B_2$ } & $\Big[x=x_2\neq 0, y=3 x_2, z=-\frac{3}{2}, v= 0, k=\frac{1}{2}\Big]$ & $0$& $\frac{1}{2}$ \\
 \hline   
    
   \parbox[c][1.3cm]{1.3cm}{$B_3$ } &  $\big[x=\frac{1}{8}, y=\frac{1}{4}, z=-1, v=0, k^2-k\neq 0 \big]$ & $-\frac{1}{3}$& $0$ \\

   \hline
   \parbox[c][1.3cm]{1.3cm}{$B_4$} &  $\big[ x=x_4\ne 0, 
   y=\frac{1}{2} (1-4 x_4),z=1-\frac{1}{4 x_4}, v=0, k=\frac{1}{2}\big]$ & $-1+\frac{1-4 x_4}{6 x_4}$& $-1+\frac{1-4 x_4}{4 x_4}$ \\
\hline
   
 \parbox[c][1.3cm]{1.3cm}{$B_5$} &  $\big[x=x_5\neq 0, y=0,  z=0, v=0, k=\frac{4 x_5}{4 x_5+1}, 4 x_5+1\neq 0 \big]$ & $-1$& $-1$ \\
 \hline
    \end{tabular}
    \caption{Critical points and corresponding values of $\omega_{tot}$, $q$ for  Model-\ref{Model-II}.}
    \label{modelIIcriticalpoints}
\end{table}
\begin{table}[H]
    \centering 
    \begin{tabular}{|c |c |c |c|} 
    \hline\hline 
    \parbox[c][0.9cm]{1.3cm}{\textbf{Critical Points}
    }& \textbf{Eigenvalues} & \textbf{Stability}& \textbf{Acceleration}\\ [0.5ex] 
    \hline\hline 
    
    \parbox[c][1.3cm]{1.3cm}{$B_1$ } & $\left[0,1,\frac{-\sqrt{4 x_1-47 x_1^2}-x_1}{2 x_1},\frac{\sqrt{4 x_1-47 x_1^2}-x_1}{2 x_1}\right]$ & Unstable & Never \\
    \hline
    \parbox[c][1.3cm]{1.3cm}{$B_2$ } & $\left[0,-1,\frac{-\sqrt{8 x_2-71 x_2^2}-3 x_2}{4 x_2},\frac{\sqrt{8 x_2-71 x_2^2}-3 x_2}{4 x_2}\right]$ & $\frac{1}{10}<x_2\leq \frac{8}{71}$ & Never \\
    \hline
   \parbox[c][1.3cm]{1.3cm}{$B_3$ } &  $\left[-2,-2,-1,\frac{2 \left(2 k-1\right)}{k-1}\right]$ & Stable for $\frac{1}{2}<k<1$ & Never\\
   \hline
   \parbox[c][1.3cm]{1.3cm}{$B_4$ } &  $\left[0,-\frac{12 x_4-1}{2 x_4},-\frac{10 x_4^6-x_4^5}{2 x_4^6},-\frac{16 x_4^6-x_4^5}{4 x_4^6}\right]$ &Stable for $x_4<0\lor x_4>\frac{1}{10}$ &  $x_4<0\lor x_4>\frac{1}{8}$ \\
   \hline
   \parbox[c][1.3cm]{1.3cm}{$B_5$ } &  $\left[-4,-\frac{3 \left(k^2-k\right)}{\left(k-1\right) k},\frac{-3 k^2+3 k-\mu_1}{2 \left(k-1\right) k},\frac{-3 k^2+3 k+\mu_1}{2 \left(k-1\right) k}\right]$ & Stable for $\frac{8}{25}\leq k<\frac{1}{2}$ & Always \\
 \hline
    \end{tabular}
    \caption{Eigenvalues corresponding to the critical points with stability and acceleration conditions for Model-\ref{Model-II}.\\ ( $\mu_1=\sqrt{25 k^4-58 k^3+41 k^2-8 k}$ .)}
    \label{modelIIeigenvalues}
\end{table}
\begin{table}[H]
    \centering 
    \begin{tabular}{|c |c |c |c|} 
    \hline\hline 
    \parbox[c][0.9cm]{1.3cm}{\textbf{Critical Points}
    }& $\Omega_r$ & $\Omega_m$&$\Omega_{DE}$\\ [0.5ex] 
    \hline\hline 
    \parbox[c][1.3cm]{1.3cm}{$B_1$ } & $1-12 x_1$ & $0$& $12 x_1$ \\
    \hline
    \parbox[c][1.3cm]{1.3cm}{$B_2$ } & $0$ & $1-10 x_2$ &$10 x_2$\\
    \hline
   \parbox[c][1.3cm]{1.3cm}{$B_3$ } &  $0$ & 0 & 1 \\
   \hline
   \parbox[c][1.3cm]{1.3cm}{$B_4$ } &  $0$ & $0$ & $1$ \\
   \hline
   \parbox[c][1.3cm]{1.3cm}{$B_5$ } &  $0$ &  $0$ &$1$ \\
 \hline
    \end{tabular}
    \caption{Standard density parameters corresponding to the critical point for Model-\ref{Model-II}.}
    \label{densityparametersm-2}
\end{table}

\begin{figure}[H]
    \centering
    \includegraphics[width=65mm]{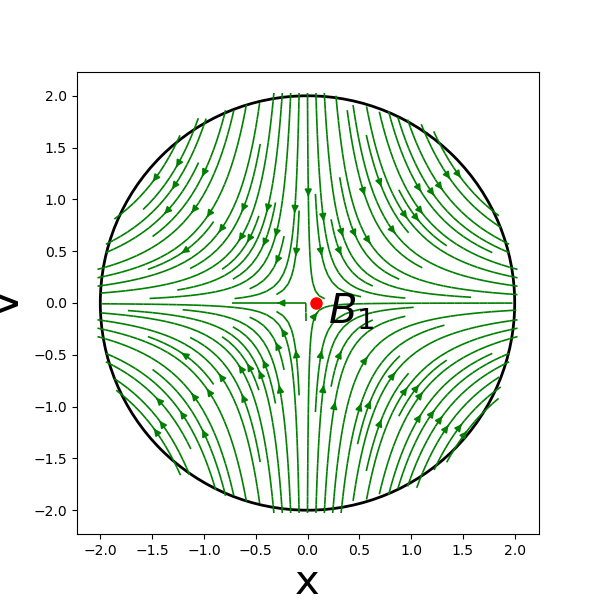}
    \includegraphics[width=65mm]{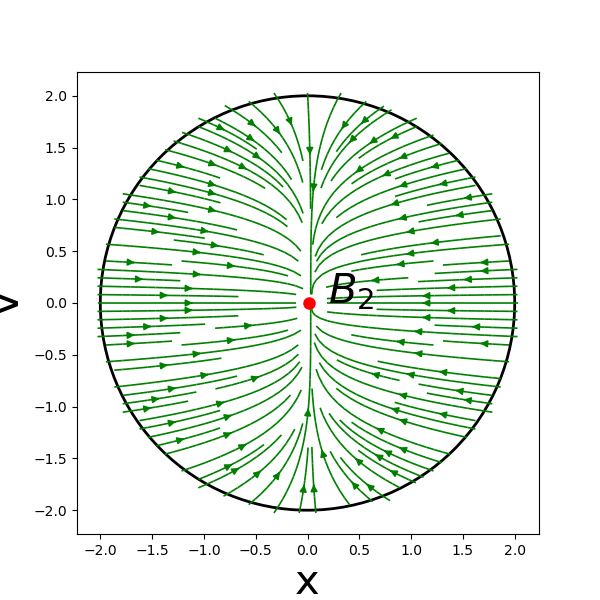}
    \includegraphics[width=65mm]{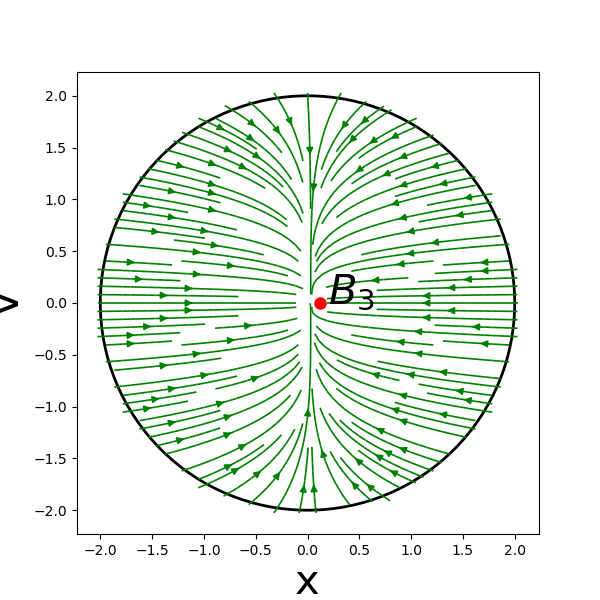}
    \includegraphics[width=65mm]{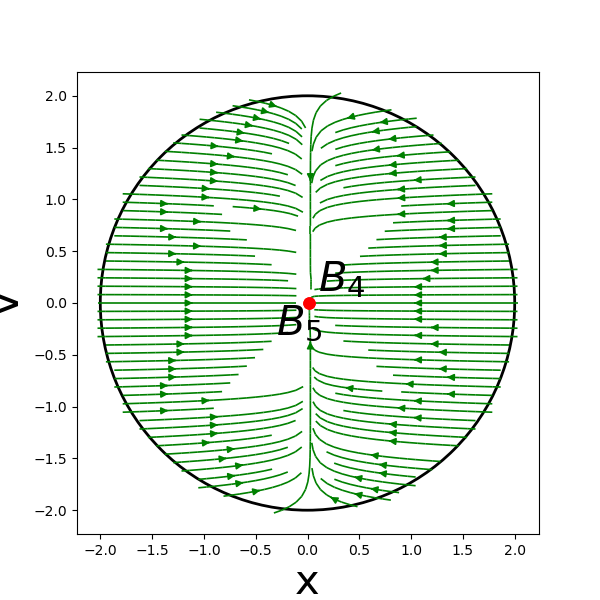}
    \caption{ $2D$ phase portrait for the dynamical system with $k=0.785$ ( Model-\ref{Model-II}). } \label{phasespacem2}
\end{figure}

\begin{figure}[H]
    \centering
    \includegraphics[width=75mm]{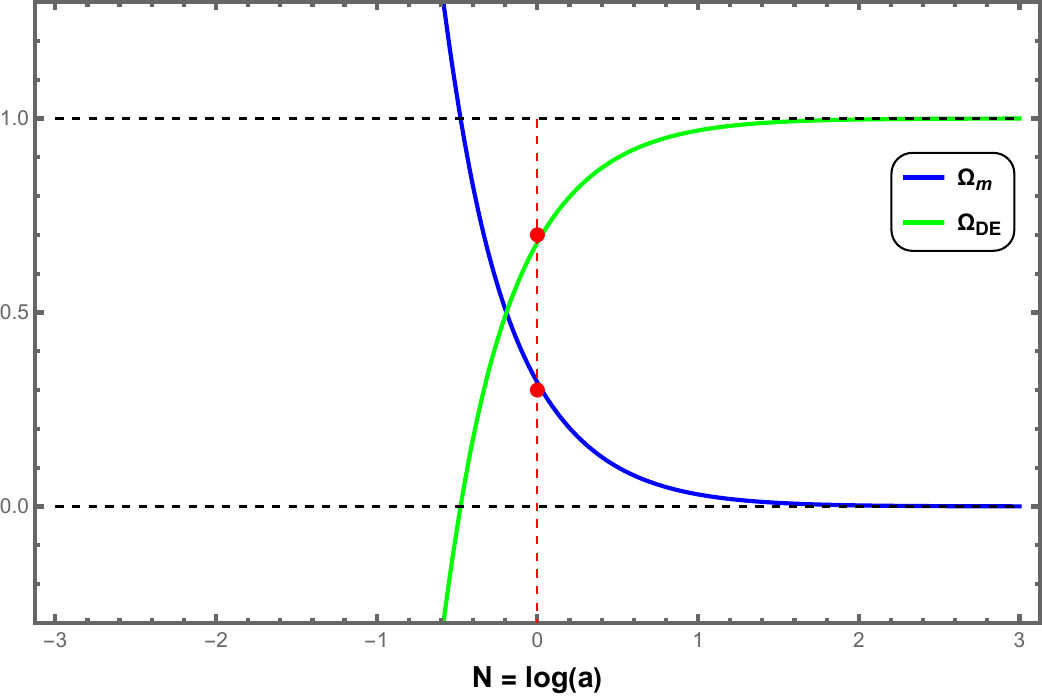}
    \includegraphics[width=75mm]{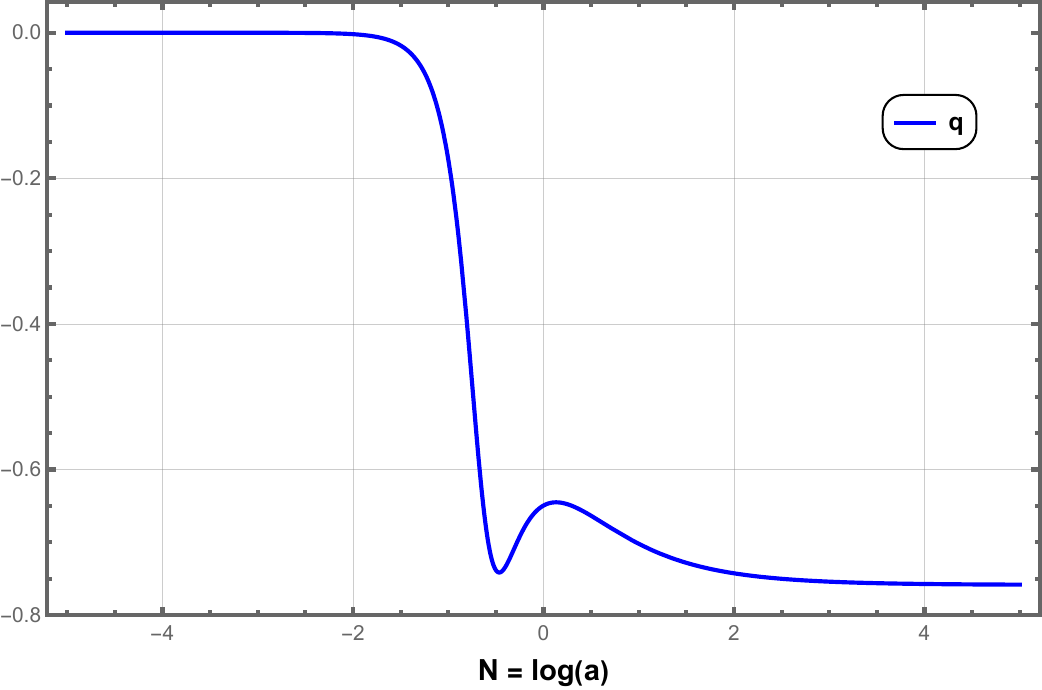}
    \caption{Behaviour of density parameters for matter and DE, $q$ in terms of the redshift for the dynamical system with $k=0.785$ ( Model-\ref{Model-II}). } \label{evolutionm2}
\end{figure}

Next, we shall provide detailed discussions on the behaviour of the critical points obtained:

\begin{itemize}
\item \textbf{Radiation-Dominated Critical Point:} The critical point  $B_1$ represents the non-standard radiation-dominated era of the evolution of the Universe. This critical point will represent the standard radiation-dominated era at $x_1=0$, which can be analyzed from Table \ref{densityparametersm-2}. The presence of eigenvalues $1, 0$ confirms this critical point is unstable and non-hyperbolic in nature. The phase space trajectories can be analyzed through Fig. \ref{phasespacem2}, and the unstable nature of the critical point can be confirmed. As presented in Table \ref{modelIIcriticalpoints}, the condition on the model parameter $k=\frac{1}{2}$ is applicable to the critical points starting from $B_1$ to the critical points representing different epochs of Universe evolution, i.e., $B_2$ and $B_4$.

\item \textbf{Matter-Dominated Critical Point:} At the critical point $B_2$, the value of $\omega_{tot}=0$ and $q=\frac{1}{2}$, hence it confirms that the critical point is of the matter-dominated era. This critical point represents a standard matter-dominated era at $x_2=0$ where $\Omega_{m}=1$ can be visualized from Table \ref{densityparametersm-2}. The presence of eigenvalue 0 implies that this critical point is normally hyperbolic and is showing stable behavior at $\frac{1}{10} < x_2 \le \frac{8}{71}$ where the other eigenvalues lie in the negative region. The phase space trajectories in Fig. \ref{phasespacem2} show that this is a matter-dominated attractor solution.

\item \textbf{Transition Critical Point :}
As we are aware, accelerating behaviour can be obtained at $q<0$ and $\omega <-\frac{1}{3}$. At the critical point $B_3$, we have $q=0$ and $\omega =-\frac{1}{3}$ refer Table \ref{modelIIcriticalpoints}. Hence, this critical point can not describe the accelerating expansion. This critical point shows its existence at $k^2 \ne k$ and stable at $\frac{1}{2} < k<1$. The attracting nature of the phase space trajectories can be studied from Fig. \ref{phasespacem2}. But due to the choice of dynamical variable $Z=\frac{\dot{H}}{H^2}$, the deceleration parameter will only show dependency on the variable $Z$, and may be this is the cause the deceleration parameter fails to produce transition phase, and the same can be observed in Fig. \ref{evolutionm2}.

\item \textbf{DE-Dominated Critical Points :}
The critical point $B_4$ in which the $\omega_{tot}, q$ depends on the dimensionless variable $x$ and is capable of explaining both the early phases and the late time of cosmic expansion depending upon the choice of the value for variable $x$. From Table \ref{modelIIeigenvalues}, one can analyze the eigenvalues of the Jacobean matrix at this critical point for the system presented in Eq. (\ref{dynamicalsystemmodel2}). At this critical point, eigenvalues contain zero and hence are non-hyperbolic. Since the number of vanishing eigenvalues is equal to the dimension of the set of eigenvalues, so it is normally hyperbolic and is stable at $x<0$ or $x>\frac{1}{10}$. The acceleration of the Universe can be described at $x<0$ or $x>\frac{1}{8}$, and the same can be seen in Table \ref{modelIIeigenvalues}. This critical point is an attractor, and the attracting nature of the phase space trajectories can be studied in Fig. \ref{phasespacem2}.\\

The critical point $B_5$ is a de-sitter solution with $\omega_{tot}=q=-1$. This critical point show stability at $\frac{8}{25} \le k \le \frac{1}{2}$. This is also an attractor solution, and the phase space diagram is presented in Fig. \ref{phasespacem2}. One thing to be noted is that these critical points represent the standard DE-dominated era of Universe evolution with $\Omega_{DE}=1$ and are presented in Table \ref{densityparametersm-2}.
\end{itemize}

The behavior of standard density parameters for matter and DE have been plotted in Fig. \ref{evolutionm2}. These parameters take the value $\Omega_m \approx 0.3$ and $\Omega_{DE} \approx 0.7$ at the present time and show compatibility to the present observational studies \cite{Aghanim:2018eyx,Kowalski_2008}. The deceleration parameter lies in the negative region and is showing the agreement in describing the late time accelerating behaviour.
\section{Results and Conclusion}\label{Conclusion}

This work presents a systematic dynamical system analysis for two cosmological viable models in the higher order teleparallel $f(T, B, T_G, B_G)$ gravity formalism. The dynamical system analysis approach plays a crucial role in the selection of a model. As discussed in \cite{bohmer2017dynamicalsystem}, the standard model of cosmology can be
highlighted by the sequence of the following cosmological evolution eras:
Inflation $\rightarrow$ radiation-dominated era $\rightarrow$ matter-dominated era $\rightarrow$ DE-dominated era and a viable cosmological model have to be able to retrace at least parts of this sequence. In this study, we have demonstrated that both models we have studied are well-motivated and capable of retracing the mentioned sequence.

The Model--\ref{Model-I} describes a radiation-dominated era at unstable critical points $A_1, A_2$ followed by the critical point $A_3$ which represents a matter-dominated era can be analyzed using Tables \ref{modelIcriticalpoints},\ref{modelIeigenvalues}. The stable behavior of DE solutions $A_4$ and $A_5$ have been obtained. Moreover, the critical point $A_5$ represents the de-Sitter solution; both are attractors, and the same can be visualized using Fig. \ref{phasespacem1}. For the critical point $A_4$, since the value of $\omega_{tot}, q$ depends upon free parameter Z,  it shows its ability to describe both the early and the late time epochs of the Universe's evolution. For Model-\ref{Model-I}, the 2-D phase space, the plots of standard density parameters for matter and DE, and the deceleration parameter have been presented in Figs. \ref{phasespacem1}-\ref{evolutionm1}. The results obtained from these figures are compatible with the current observational studies \cite{Aghanim:2018eyx,Kowalski_2008}.

Similar to the Model-\ref{Model-I}, Model-\ref{Model-II} generates the sequence of radiation, matter, and DE attractor solutions. The critical point $B_1$ is the only point representing the radiation-dominated era in Model-\ref{Model-II}. The key difference in both the models is an additional critical point describing the transition phase (critical point $B_3$) is present in Model-\ref{Model-II}.  In the matter-dominated solution, the critical point $B_4$ is normally hyperbolic and is showing stability at $x_4 < 0$ or $x_4>\frac{1}{10}$. The de-Sitter solution in Model-\ref{Model-II} represents standard DE era and is stable at $\frac{8}{25} \le k \le \frac{1}{2}$. The model parameter $k$ takes the value $\frac{1}{2}$ at which critical points represent the radiation, matter, and DE eras of Universe evolution. The plots, in this case, presented in Figs.\ref{phasespacem2} and \ref{evolutionm2} are compatible with the observational studies \cite{Aghanim:2018eyx,Kowalski_2008,Capozziellomnras} at the present time.

This study extends the analysis made in \cite{Franco:2020lxx,
KADAM2024Aop} and combine the results in a more general way to test the cosmological applications of the presented models. Moreover, this guarantees to move forward to check the observational outcomes of these models. 

\section*{Acknowledgements}
SAK acknowledges the financial support provided by the University Grants Commission (UGC) through Senior Research Fellowship (UGC Ref. No.: 191620205335), and BM acknowledges IUCAA, Pune, India, for hospitality and support during an academic visit where a part of this work has been accomplished. 
\section*{References} 
\bibliographystyle{utphys}
\bibliography{references}

\end{document}